\documentclass[%
 reprint,
 superscriptaddress,
 amsmath,amssymb,
 aps,
]{revtex4-2}
\usepackage{graphicx}

\usepackage[version = 4]{mhchem} 
\usepackage{xcolor}
\usepackage{soul}
    \sethlcolor{yellow}

\newcommand{\dd}{\mathrm{d}}

\newcommand{\eref}[1]{Eq.~\ref{#1}}
\newcommand{\fref}[1]{Fig.~\ref{#1}}

\graphicspath{ {./figures/} }

\begin{document}

\title{Optical Detection of Coefficient of Thermal Expansion via Frequency Tracking at Cryogenic Temperatures}

\author{Yanli Shi}
  \affiliation{Shandong Institute of Advanced Technology, Jinan, Shandong 251000, China}
\author{Hui Wu}
  \affiliation{Shandong Institute of Advanced Technology, Jinan, Shandong 251000, China}
\author{Guangyu Li}
  \affiliation{Institute of Thermal Science and Technology, Shandong University, 250061, Jinan, China}
\author{Qun Cao}
  \email{qun.cao@iat.cn}
  \affiliation{Shandong Institute of Advanced Technology, Jinan, Shandong 251000, China}
\author{Zhe He}
  \email{zhe.he@iat.cn}
  \affiliation{Shandong Institute of Advanced Technology, Jinan, Shandong 251000, China}
\author{Zheng Cui}
  \affiliation{Shandong Institute of Advanced Technology, Jinan, Shandong 251000, China}


\begin{abstract}
Accurate characterization of the coefficient of thermal expansion~(CTE) in nanoscale thin films is critical for predicting residual stress and ensuring the reliability of modern micro-electromechanical systems and advanced semiconductor architectures.
The primary limitation is that traditional methods depend on precise thickness measurements or are restricted to crystalline materials.
In this work, we present an \textit{in-situ}, non-destructive optical methodology to determine the CTE of suspended nanofilms by tracking the temperature-dependence of their vibrational frequency.
This approach eliminates thickness measurements entirely and works for amorphous films where X-ray diffraction~(XRD) fails.
We experimentally validate this technique across a broad temperature range, evaluating membranes in their native microstructure state without requiring additional destructive patterning or complex sample preparation.
This non-contact optical method achieves uncertainty below $5.3\times10^{-8}$~K\textsuperscript{-1}, substantially improving upon conventional thin-film CTE metrology.
\end{abstract}

\maketitle





\section{Introduction}

Thermal expansion mismatch drives issues in modern nanoscale devices.
Stress-induced deformation degrades quantum processor coherence, shifts MEMS resonator frequencies out of specification, and fractures semiconductor interconnects during thermal cycling~\cite{Han23Jiang,Agonafer08Pekin}.
These issues stem from coefficient of thermal expansion~(CTE), a fundamental thermomechanical property that governs the dimensional stability of materials across varying thermal environments.
As integration density increases and operating environments span from cryogenic quantum systems to high-temperature power electronics, precise CTE characterization becomes essential for predicting residual stress, preventing mechanical fatigue, and ensuring device reliability.

Among materials widely used in cryogenic and room-temperature nanoscale platforms, silicon nitride (\ce{Si3N4}) plays an important role.
Renowned for its exceptional mechanical toughness, chemical inertness, and broad optical transparency, thin-film \ce{Si3N4} has become the cornerstone material for high-performance MEMS, optomechanical resonators~\cite{Serra16Bonaldi,Zwickl08Harris}.
These films also serve as passivation layers in CMOS fabrication~\cite{Blanc13Joubert,Ciftlik11Gijs}, ultra-thin membranes for liquid-cell transmission electron microscopy~(TEM)~\cite{Pu20Robertson} and solid-state nanopores~\cite{Chou20Drndic,Dekker07}.
However, \mbox{\ce{Si3N4}} nanofilms exhibit highly variable CTE values that depend sensitively on deposition temperature, precursor gas ratios, and chamber pressure~\cite{Martyniuk06Faraone,Martyniuk06THESIS}.
Consequently, the CTE of the \ce{Si3N4} film is not a universal constant, but a highly variable and critical parameter.
When bonded to substrates with mismatched thermal properties, even small CTE uncertainties translate to severe differential stress during thermal cycling.
Precise, temperature-resolved CTE measurements for \ce{Si3N4} nanofilms are essential for understanding thermal expansion behavior across temperature ranges, guiding material selection for cryogenic applications, and predicting thermomechanical stability in quantum devices and space instrumentation.
Yet existing CTE measurement protocols remain insufficient for nanoscale thin films.
Conventional techniques either require precise thickness measurements, are restricted to crystalline materials, or lack the sensitivity required for sub-micron structures across cryogenic to room temperature ranges.

Established methods for measuring CTE of a nanofilm over cryogenic-to-room-temperature ranges include X-ray diffraction~(XRD)~\cite{James01Evans,Wolff99,Liao18Goorsky,Lindsay-Scott07Dobson}, optical imaging~\cite{James01Evans,Bing09Asundi,Dudescu06Nebel}, and substrate-curvature method~\cite{Martyniuk06Faraone,Thurn04Hughey}.
XRD determines CTE by tracking lattice parameter changes from Bragg diffraction peaks.
While nondestructive, XRD suffers from overwhelming substrate signal due to the large substrate-to-film thickness ratio, and CVD ceramic films are predominantly amorphous or polycrystalline~\cite{Pierson99}, yielding weak or absent diffraction peaks. 
Furthermore, the XRD results do not necessarily correspond to the macroscopic CTE of the material, especially when the sample contains multiple phases~\cite{James01Evans}.
Tracking the area change of a sample at different temperatures by taking high-resolution optical images can also determine the CTE of nanostructures.
However, nanofilms are grown on top of a relatively much thicker substrate.
As a consequence, the area change of the nanofilm reflects essentially the thermal expansion of the substrate, not the nanofilm.
Consequently, area-based optical imaging cannot isolate the nanofilm CTE while the film remains on its native substrate.
In contrast, by taking advantage of the substrate-nanofilm structure, the curvature of the system can be used to measure the residual stress of the nanofilm through the Stoney formula~\cite{Stoney09}.
The residual stress can also be used to determine the CTE of the nanofilm with the knowledge of the biaxial modulus.
However, in addition to biaxial modulus, the curvature method requires the thickness of both the substrate and the nanofilm, which are major sources of the uncertainty in the result.
Moreover, the substrate and nanofilm thickness uniformity substantially affects the accuracy of the resulting CTE profile via the substrate-curvature method.
Collectively, existing methods either fail for amorphous films (XRD), cannot isolate film properties from substrate contributions (optical imaging), or propagate large thickness-measurement uncertainties (curvature), limiting their applicability to CVD nanofilms on native substrates.


To overcome these limitations of current metrology, this work presents a high-precision, robust, non-invasive \textit{in-situ} optical method for determining the CTE of suspended nanofilms via resonant frequency tracking.
Unlike geometric methods that measure area changes, our frequency-based approach directly couples to the film's intrinsic thermal properties through its vibrational modes.
A primary advantage of this approach is its ability to evaluate the suspended nanofilms in their native microstructure and stress state, without additional destructive patterning or secondary material modification.
By eliminating the need for independent thickness measurements of the film and substrate in the CTE extraction, we remove a major driver of error that plagues traditional indirect methods.
Validated from $80$~K to $320$~K on suspended \ce{Si3N4} nanofilms, this approach yields reproducible, temperature-resolved CTE profiles with uncertainties substantially below those of conventional approaches.

\section{Theory}

The vibrational frequency for a square membrane under uniform in-plane stress is given by
\begin{align}
  f_{(m,n)}=\frac{\sqrt{m^2+n^2}}{2l}\sqrt{\frac{\sigma}{\rho}}
  \label{eq:eigenfrequency}
  ,
\end{align}
where $l$ is the lateral size, $\rho$ is the mass density, $\sigma$ is the stress of the membrane, $m,n=1,2,3,\cdots$ are the mode indices of the vibration.
Assuming an isotropic membrane subject to uniform in-plane tension, the stress $\sigma$ is described by the Hooke's law
\begin{align}
  \sigma=\frac{E}{1-\nu}\epsilon_\text{el}
  ,
\end{align}
where $E$ is the Young's modulus, $\nu$ is the Poisson ratio, $\epsilon_\text{el}$ is the elastic strain of the membrane.
Because the membrane is deposited on top of the substrate, its total strain $\epsilon$, comprising its thermal strain $\epsilon_\text{th}$, elastic strain $\epsilon_\text{el}$ and intrinsic strain $\epsilon_\text{in}$, must equal the thermal strain of the substrate $\epsilon_{s,\text{th}}$
\begin{align}
  \epsilon=\epsilon_\text{th}+\epsilon_\text{el}+\epsilon_\text{in}=\epsilon_{s,\text{th}}
  .
\end{align}
Consequently, the temperature-dependent elastic strain can be expressed as the integral of the differential thermal expansion
\begin{align}
  \epsilon_\text{el}(T)
  &=\epsilon_{s,\text{th}}(T)-\epsilon_\text{th}(T)-\epsilon_\text{in}
  \nonumber\\
  &=\int_{T_0}^{T}[\alpha_s(T')-\alpha(T')]\dd T'
  -\epsilon_\text{in}
  ,
\end{align}
where $T_0$ is the deposition temperature and $\alpha$ and $\alpha_s$ are the CTE of the membrane and substrate, respectively.
By substituting the elastic strain into Hooke's law, and subsequently into \eref{eq:eigenfrequency} for the fundamental mode ($m=n=1$), we can differentiate with respect to temperature to explicitly isolate the CTE of the membrane
\begin{align}
  \alpha(T)
  =
  \alpha_s(T)
  -
  2\rho l^2\frac{1-\nu}{E}\frac{\partial f_{(1,1)}^2(T)}{\partial T}
  \label{eq:CTE_temperature_dependence}
  .
\end{align}
\eref{eq:CTE_temperature_dependence} demonstrates that the CTE profile of the membrane can be determined simply by tracking the temperature-dependent shift of its fundamental eigenfrequency.
Note that neither the membrane nor the substrate thickness enters the formula, removing a major source of uncertainty.
Furthermore, the temperature-independent intrinsic strain $\epsilon_\text{in}$ contributes only a constant offset to the eigenfrequency, which drops out of the temperature derivative in \eref{eq:CTE_temperature_dependence} and therefore has no effect on the extracted CTE.
Because this eigenfrequency, typically in the kHz or MHz regime, can be measured non-destructively and with exceptional precision using optical techniques, this approach significantly reduces the overall uncertainty of the CTE measurement.
The exceptional precision of optical interferometric detection enables sub-Hz frequency resolution.
In our measurements, the typical frequency uncertainty is about $5$~Hz for resonant frequencies on the order of $100$~kHz, corresponding to a relative precision of $5\times10^{-5}$. 
This high frequency resolution translates directly to high CTE precision through \eref{eq:CTE_temperature_dependence}, as the measurement uncertainty is dominated by material property uncertainties (E, $\nu$) rather than frequency measurement noise. 

Note that \eref{eq:CTE_temperature_dependence} treats the Young's modulus $E$ and Poisson's ratio $\nu$ as temperature-independent.
This simplification is justified, as both parameters exhibit negligible variation over the evaluated temperature range~\cite{Chuang04Ghodssi,Martyniuk06THESIS}. 
The assumption of constant lateral size $l$ and mass density $\rho$ are rigorously estimated in appendix.


\section{Experimental Setup}

\begin{figure*}[htb!]
  \centering
  \includegraphics[width=0.8\linewidth]{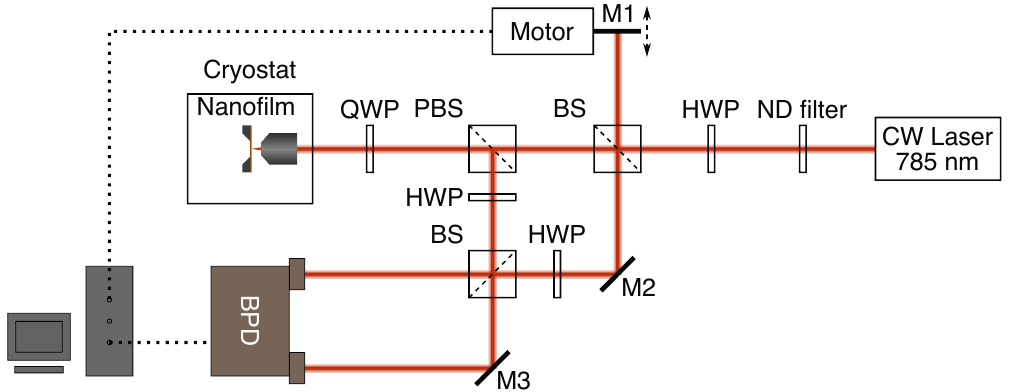}
  \caption{\label{fig:exp_setup}
    Schematic of the Mach-Zehnder interferometer for measuring the temperature-dependent resonant frequency of suspended nanofilms.
    The laser beam is split into two arms: the signal arm passes through the cryostat containing the nanofilm, while the reference arm is reflected from mirror M2.
    A motor-controlled mirror M1 with PID feedback (dashed box) actively stabilizes the interferometer at maximum sensitivity.
    The balanced photodetector~(BPD) measures the interference signal, which is processed by a computer to extract the power spectral density and track the fundamental eigenfrequency.
    CW, continuous-wave; ND, neutral density filter; HWP, half-wave plate; QWP, quarter-wave plate; BS, beam splitter; PBS, polarizing beam splitter; M, mirror.
  }
\end{figure*}

The experimental apparatus, illustrated in \fref{fig:exp_setup}, employs a Mach-Zehnder interferometer with balanced photodetection to measure the thermally-driven vibrational noise of suspended \ce{Si3N4} nanofilms across a broad temperature range.
A continuous-wave $785$~nm laser with incident power of $5$~$\mu$W (attenuated by a neutral density~(ND) filter to minimize heating effects) and polarization-controlled by a half-wave plate~(HWP), is split by a beam splitter~(BS) into two interferometer arms.
The upper arm contains a motor-controlled mirror M1 for active stabilization, followed by a polarizing beam splitter~(PBS) and quarter-wave plate~(QWP) that enable double-pass transmission through the cryostat.
The suspended nanofilm under investigation is mounted on a cold finger inside a liquid nitrogen cryostat, which provides temperature control with stability better than $1$~mK from $80$~K to $320$~K.
The nanofilm is fixed to an oxygen-free (TU0) copper plate with a central aperture that allows the laser beam to pass through, facilitating optical alignment.
The copper plate is thermally anchored to the cold finger using indium (\ce{In}) foil to ensure efficient thermal contact.
The lower reference arm directs the beam through polarization optics~(HWPs) to mirrors M2 and M3.
Both arms recombine at a second BS, and the interfering beams are detected by a balanced photodetector~(BPD).
The BPD output is processed by a computer that extracts the power spectral density~(PSD) of the nanofilm's motion and implements a software-based PID controller to continuously adjust M1, maintaining the interferometer at maximum sensitivity by minimizing the balanced signal.
The fundamental resonant frequency $f_{(1,1)}$ is tracked as a function of temperature, enabling determination of the nanofilm's CTE profile according to \eref{eq:CTE_temperature_dependence}.

\section{Results}\label{sec:results}

The experimental samples consist of three commercially available, suspended low-stress \ce{Si3N4} nanofilms (sample A, Norcada Inc. NX5100D; sample B, NX5150D; and sample C, NX5200D nanofilms).
Each nanofilm has a nominal thickness of 200 nm and is supported by a $5\times5\times0.2$~mm \ce{Si} substrate.
The etched windows feature different nominal lateral dimensions ranging from $l=1$~mm to $2$~mm with fundamental resonant frequencies spanning from $\sim25$~kHz to $\sim100$~kHz (factor of 4 variation). 
We performed X-ray diffraction~(XRD) measurements on these nanofilms.
The diffraction signal is dominated by the \mbox{\ce{Si}} substrate, and the predominantly amorphous \mbox{\ce{Si3N4}} layer yields weak or absent peaks, so the XRD data fails to estimate the CTE of the nanofilm.
The membranes also exhibit different initial stress states, as evidenced by frequency offsets between thermal cycles, and originate from different production batches.

Temperature-dependent measurements were initiated at $80$~K, with the cryostat heated at a controlled sweep rate not exceeding $0.27~\text{K}\cdot\text{s}\textsuperscript{-1}$.
To ensure thermal equilibrium, the temperature was stabilized at predefined setpoints for $5$~min, which is sufficient for full equilibration as verified in our previous work~\cite{Li26Cui1} and confirmed by direct measurement, achieving a thermal stability of better than $1$~mK.
During each dwell period, the fundamental eigenfrequency of the membrane was sampled multiple times to extract the mean resonant frequency and its associated standard error for that specific temperature.

\begin{figure}[htb!]
  \centering
  \includegraphics[width=\linewidth]{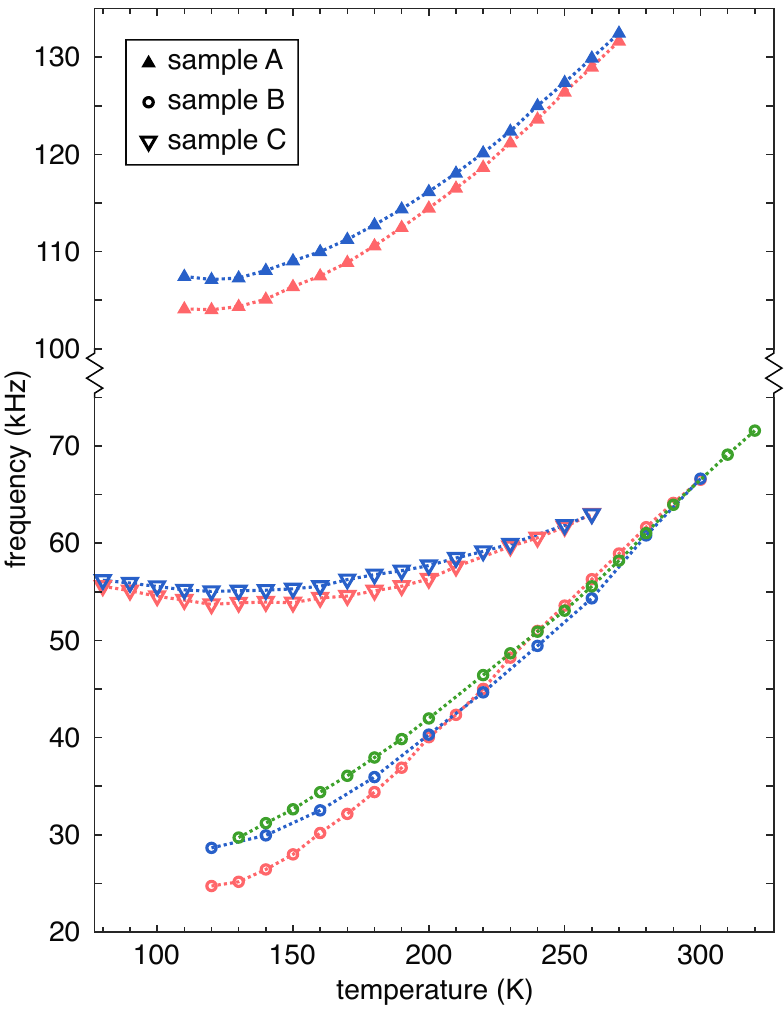}
  \caption{\label{fig:freq_vs_temp}
    Measured frequency as a function of temperature for sample A (closed upward triangle), sample B (open circle), and sample C (open downward triangle).
    Colors within same figure denote different runs of the same membrane.
    The error bar (not exceeding $40$~Hz) is too small to be displayed in the graphs.
    All panels share identical axis spans such that the trends between graphs are comparable.
  }
\end{figure}

\fref{fig:freq_vs_temp} presents the measured eigenfrequency for multiple nanofilm samples as a function of temperature.
A positive (negative) slope indicates the CTE of the \ce{Si3N4} nanofilm is smaller (larger) than that of the \ce{Si} substrate.
When the CTE of the nanofilm exactly matches the CTE of the \ce{Si} substrate, the slope of the frequency-temperature curve becomes zero, which is observed for samples A and C near $130$~K.
This produces a minimum in the $f(T)$ curve where the frequency first decreases with temperature, reaches a minimum, then increases.
Slight absolute frequency offsets between runs of the same nanofilm likely originate from the relaxation of residual stress during thermal cycling.
Frequency differencies among three samples come from various laterial sizes and intrinsic stress states.
Our method is robust against such frequency drift, as it relies fundamentally on the slope of $f(T)$ of the eigenfrequency.

\begin{figure}[htb!]
  \centering
  \includegraphics[width=\linewidth]{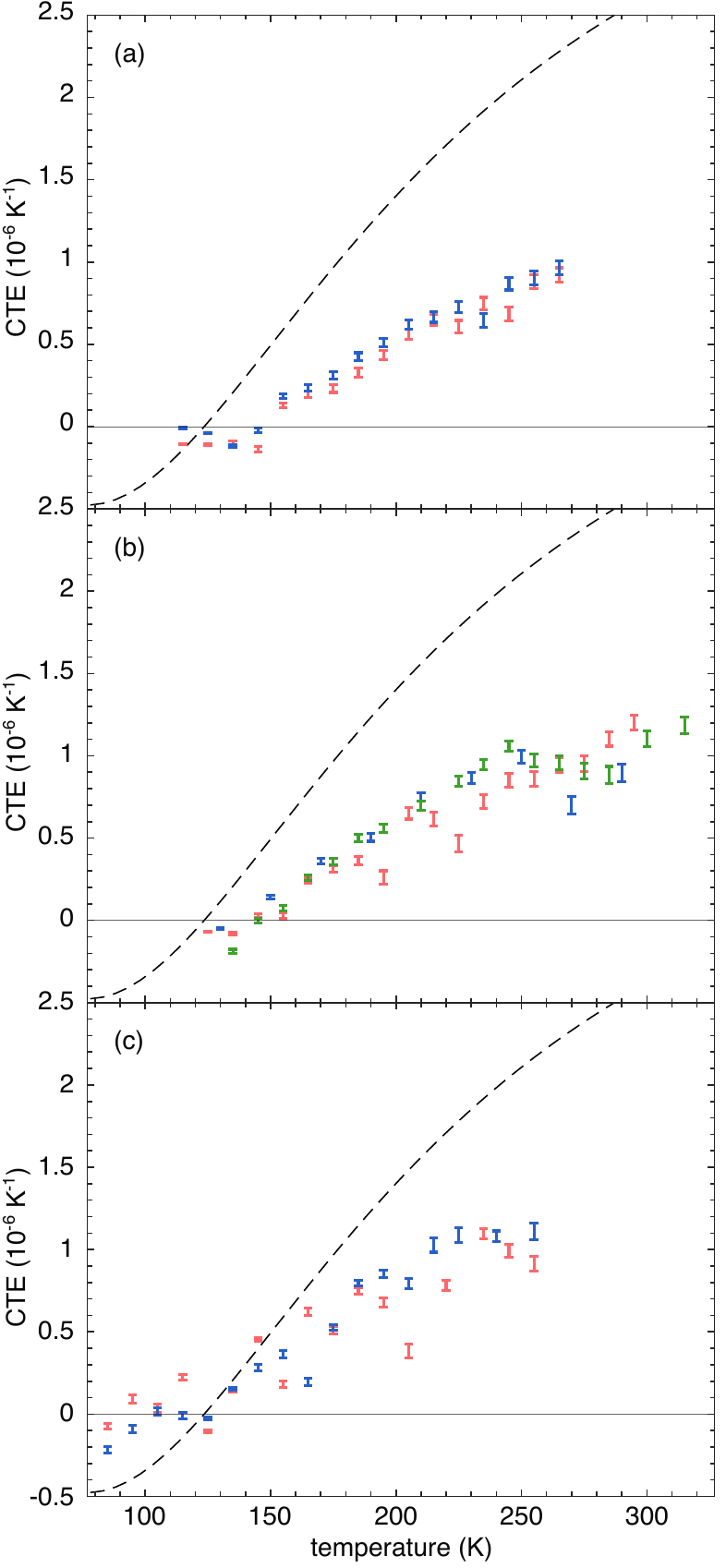}
  \caption{\label{fig:results}
    Calculated CTE of (a) sample A, (b) sample B, and (c) sample C.
    Colors within same figure denote different runs of the same membrane.
    Black dashed line indicates the CTE of the \ce{Si} substrate~\cite{Middelmann15Schodel}.
  }
\end{figure}

The CTE value $\alpha(T)$ was evaluated using \eref{eq:CTE_FD} for each data run.
Error bars in \fref{fig:results} represent the propagated uncertainty from \eref{eq:CTE_uncertainty}.
The temperature dependence of the eigenfrequency reflects the CTE of the nanofilm based on \eref{eq:CTE_temperature_dependence}.
For calculations in this work, we adopted $E=263\pm5$~GPa~\cite{Chuang04Ghodssi} and $\nu=0.23\pm0.02$~\cite{Edwards04Sharpe} for Young's modulus and Poisson ratio, respectively.
The lateral sizes of the nanofilms were measured to be $l_\text{A}=1.026$~mm, $l_\text{B}=1.537$~mm and $l_\text{C}=2.029$~mm by optical microscopy, with uncertainty of $1$~$\mu$m.
The mass density of the \ce{Si3N4} is taken as $\rho=3.17$~{g$\cdot$cm\textsuperscript{-3}}~\cite{Haynes17}. 
The CTE of the \ce{Si} substrate refers to Ref.~\cite{Middelmann15Schodel}, which offers an uncertainty on the order of $10^{-9}$~K\textsuperscript{-1} across the temperature range from $8.15$~K to $293.15$~K.
For temperature beyond this range, the CTE values were extrapolated using fifth-order polynomial fitting.
The uncertainty of CTE of the \ce{Si} substrate remains below $1.3\times10^{-8}$~K\textsuperscript{-1} for the extrapolated data.

The mass density of the \ce{Si3N4} nanofilm could be $20\%$ smaller than the stoichiometric value presented above~\cite{Hirai81Goto,Chu67Gruber,Bontempi01Pirri}.
However, this deviation will only result in a bias of the overall CTE estimation.
The actual mass density of the nanofilm can be determined using X-ray reflectometry~(XRR)~\cite{Zhou95Chen,Li98Kojima,Santucci01Lozzi,Tiilikainen08Lipsanen}. 

The extracted CTEs of three \ce{Si3N4} nanofilm samples as a function of temperature are shown in \fref{fig:results}.
For each thermal cycle, the uncertainty remains below $5.3\times10^{-8}$~K$^{-1}$ across all temperatures and samples.

The three membranes exhibit similar CTE temperature dependencies despite differences in lateral size, resonant frequency, and initial stress states.
Slight differences between the CTE curves are attributable to variations in CVD deposition parameters such as temperature, precursor gas ratios, and chamber pressure~\cite{Martyniuk06Faraone,Martyniuk06THESIS}, though the membranes likely originate from the same deposition recipe given their similar behavior.

\begin{figure}[htb!]
  \centering
  \includegraphics[width=\linewidth]{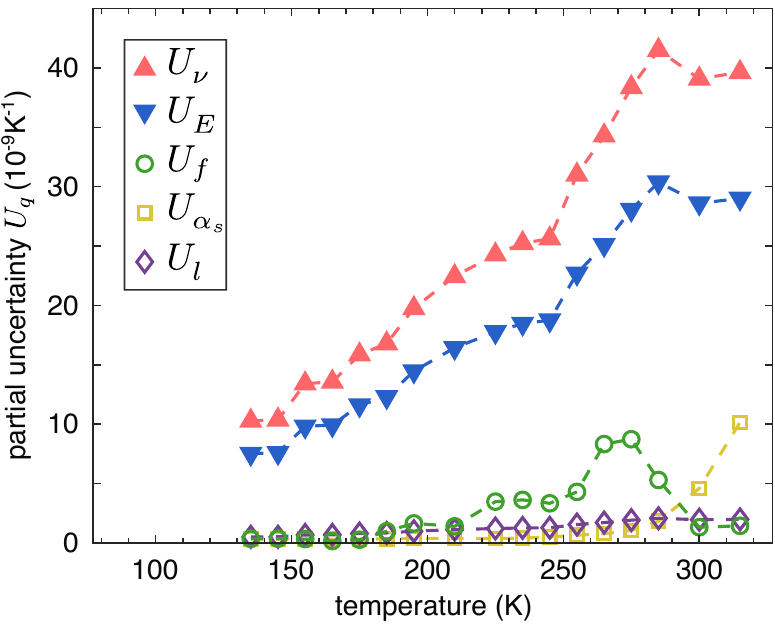}
  \caption{\label{fig:source_of_uncertainty}
    Partial uncertainty $U_q$ based on \eref{eq:CTE_uncertainty} for one run [green data points in \fref{fig:results}(b)] of sample B.
    The partial uncertainty of Poisson ratio $\nu$ (closed red upward triangle) and the Young's modulus $E$ (closed blue downward triangle) dominates the contributions.
    The partial uncertainty $U_{\alpha_s}$ (open yellow square) from the CTE of the \ce{Si} substrate rises in high temperature region ($T\gtrsim280$~K) due to extrapolation.
    The partial uncertainty $U_f$ from frequency measurement (open green circle) remains below $5\times10^{-9}$~K\textsuperscript{-1} over the most of the range, staying under $10^{-8}$~K\textsuperscript{-1} everywhere.
  }
\end{figure}

\fref{fig:source_of_uncertainty} illustrates the contributions to uncertainty in the CTE profile.
The lack of knowledge of the Poisson ratio $\nu$ and the Young's modulus $E$ accounts for the majority of the uncertainty of the calculated CTE.
Frequency and lateral-size contributions remain comparatively small.
The contribution from the substrate CTE $\alpha_s$ increases for $T\gtrsim280$~K owing to extrapolation of the silicon reference data.


As shown in \eref{eq:CTE_temperature_dependence}, the single-substrate analysis still requires the mass density $\rho$, the lateral size $l$, the Young's modulus $E$ and the Poisson ratio $\nu$ to extract the CTE profile.
However, these requirements can all be lifted for CTE extraction by adopting the double-substrate protocol~\cite{Retajczyk80Sinha,Retajczyk80Sinha1,Thurn04Hughey,Martyniuk06THESIS,Vavrille24Estevez}.
In the double-substrate protocol, nanofilms are grown on top of two different substrates [such as silicon (\ce{Si}) and sapphire (\ce{Al2O3})] with known CTE profile $\alpha_{s1}$ and $\alpha_{s2}$.
The nanofilm CTE can be determined by comparing the eigenfrequency temperature dependence of identically-sized nanofilms fabricated on substrates with different known CTEs.
\begin{align}
  \frac{E}{1-\nu}
  &=
  \frac{2\rho l^2}{\alpha_{s1}-\alpha_{s2}}\frac{\partial}{\partial T}\left(f_1^2-f_2^2\right)
  ,
  \\
  \alpha(T)
  &=
  \frac{\frac{\partial}{\partial T}\left(\alpha_{s2}f_1^2-\alpha_{s1}f_2^2\right)}{\frac{\partial}{\partial T}\left(f_1^2-f_2^2\right)}
  ,
\end{align}
where $f_i$ denotes the $(1,1)$ mode eigenfrequency of membrane on the two substrates $i=1,2$.
The double-substrate protocol eliminates the need for prior knowledge of Young's modulus, Poisson ratio and mass density.
Consequently, it can fully exploit the advantage of precise detection of our optical frequency-tracking method.

In summary, the uncertainty below $5.3\times10^{-8}$~K$^{-1}$ for individual thermal cycles and reproducible frequency measurements across multiple runs indicate the reliability of frequency-tracking CTE extraction.
The CTE profiles reported here characterize the specific low-stress \ce{Si3N4} nanofilms studied and reflect their particular CVD deposition conditions rather than universal \ce{Si3N4} constants.
The accuracy of this method can be further improved by measuring identically-sized nanofilms fabricated on substrates with different known CTEs, which would decouple film and substrate contributions without requiring independent thickness measurements.

\section{Conclusion}\label{sec:conclusion}


We have developed and experimentally validated a novel, non-destructive optical methodology for determining the CTE of suspended nanofilms.
By tracking the temperature-dependent shift of the nanofilm's fundamental resonant frequency, this technique inherently eliminates film and substrate thickness from the extraction, substantially reducing uncertainties.

Furthermore, this method offers a distinct advantage of \textit{in-situ} measurement.
As demonstrated through the characterization of commercial CVD \ce{Si3N4} nanofilms across a broad cryogenic-to-room-temperature range, the technique successfully evaluates the nanofilms in their native microstructure state.
It requires no additional destructive patterning, complex sample preparation, or secondary material modification, preserving the true macroscopic thermomechanical behavior of the film.
The optical frequency tracking yielded highly reproducible CTE profiles with uncertainties tightly constrained below $5.3\times10^{-8}$~K$^{-1}$.
Multiple thermal cycles on the same nanofilm may exhibit frequency shifts due to stress state changes, yet the extracted CTE profiles remain consistent, demonstrating that the method measures intrinsic material properties independent of initial stress conditions.
Measurements across three nanofilms with different lateral sizes ($1.0$~mm, $1.5$~mm, and $2.0$~mm) show similar CTE temperature dependencies with slight variations attributable to CVD deposition conditions, confirming the physical validity of the frequency-tracking approach.

Currently, the primary residual uncertainties in this single-substrate approach originate from the assumed material properties of the film, specifically Young's modulus and Poisson's ratio.
While these specific material-property dependencies can be eliminated altogether in the CTE extraction by future implementations of a double-substrate measurement protocol, the baseline optical method presented here already represents a substantial improvement in measurement precision over conventional techniques.
Ultimately, by providing a high-resolution, non-contact tool for cryogenic-temperature CTE characterization, this methodology delivers the metrological precision necessary to accurately model thermal stress, prevent mechanical failure, and ensure the long-term reliability of next-generation MEMS, quantum devices, cryogenic MEMS, space instrumentation, and advanced semiconductor packaging.

\section*{Acknowledgement}
This work is supported by the National Natural Science Foundation of China under grant No. 12404413, Shandong Provincial Natural Science Foundation (No. 2024HWYQ-081), and Taishan Scholars Program (No. tsqn202408307, No. tsqn202306316).

\section*{Data availability}
Data underlying the results presented in this paper can be obtained from the authors upon reasonable request.

\section*{Disclosures}
The authors declare no competing interests.


\section*{Appendix}

\subsection*{Error propagation}

Using the finite-difference method, the estimated CTE of the membrane is
\begin{align}
  \alpha\left(\frac{T_1+T_2}{2}\right)
  &=
  \alpha_s\left(\frac{T_1+T_2}{2}\right)
  \nonumber\\
  &\quad
  -
  2\rho l^2\frac{1-\nu}{E}
  \frac{f_{(1,1)}(T_2)^2-f_{(1,1)}(T_1)^2}{T_2-T_1}
  \label{eq:CTE_FD}
  .
\end{align}
When quantity $p$ is a set of combination of measurable quantities $q_i$, $p=\sum_iq_i$, $u_p$, the uncertainty of $p$, is determined by $u_{q_i}$, the uncertainty of $q_i$, through
\begin{align}
  u_p^2=\sum_i\left(\frac{\partial p}{\partial q_i}\right)^2u_{q_i}^2.
\end{align}
Therefore, the uncertainty of nanofilm CTE can be estimated by
\begin{align}
  u_\alpha^2
  &=
  u_{\alpha_s}^2+\Delta\alpha^2\bigg[\frac{4u_l^2}{l^2}+\frac{u_\nu^2}{(1-\nu)^2}+\frac{u_E^2}{E^2}
  \nonumber\\
  &\quad
  +\frac{(2f_2u_{f_2})^2+(2f_1u_{f_1})^2}{(f_2^2-f_1^2)^2}\bigg]
  \nonumber\\
  &=
  U_{\alpha_s}^2+U_{l}^2+U_{\nu}^2+U_{E}^2+U_{f}^2
  \label{eq:CTE_uncertainty}
  ,
\end{align}
where
\begin{align}
  \Delta\alpha=2\rho l^2\frac{1-\nu}{E}\frac{f_2^2-f_1^2}{T_2-T_1}
  ,
\end{align}
$u_q$ is the uncertainty of physical quantity $q$, $f_i=f_{(1,1)}(T_i)$ is the fundamental mode eigenfrequency at temperature $T_i$ and $U_q$ is the partial uncertainty of $\alpha$ propagated from $q$, e.g., for $q=l$, $U_l=2\Delta\alpha u_l/l$.

\subsection*{Temperature dependence of lateral size and mass density}

The typical value of the CTE $\alpha$ is on the order of $10^{-6}~\text{K}\textsuperscript{-1}$, resulting in a relative change of $\sim10^{-4}$ in linear dimensions of both the substrate and the nanofilm for $100$~K temperature change.
Consequently, the lateral size of the membrane $l$ varies linearly according to the CTE and the mass density $\rho$ inverse cubically.
Therefore, the relative change in the term
\begin{align}
  \rho(T) l^2(T)
  &=
  \rho l^2\left(1+\int_{T_0}^T\alpha(T)\text{d}T\right)^{-3}
  \nonumber\\
  &\quad
  \times\left(1+\int_{T_0}^T\alpha_s(T)\text{d}T\right)^2
  \nonumber\\
  &\approx
  \rho l^2\left(1+\int_{T_0}^T2\alpha_s(T)-3\alpha(T)\text{d}T\right)
  ,
\end{align}
is also on the order of $10^{-4}$, leading to a bias in $\alpha$ on the order of $10^{-10}~\text{K}\textsuperscript{-1}$ which is smaller than the uncertainty introduced by, e.g., Young's modulus.
Consequently, the change of lateral size and mass density due to temperature can be neglected.

\bibliographystyle{apsrev4-2}
\bibliography{reference}

@article{Agonafer08Pekin,
  title = {Thermo-{{Mechanical Challenges}} in {{Stacked Packaging}}},
  author = {Agonafer, Dereje and Kaisare, Abhijit and Hossain, Mohammad M. and Lee, Yongje and {Dewan-Sandur}, Bhavani P. and Dishongh, Terry and Pekin, Senol},
  year = 2008,
  journal = {Heat Transfer Engineering},
  volume = {29},
  number = {2},
  pages = {134--148},
  publisher = {Taylor \& Francis},
  issn = {0145-7632},
  doi = {10.1080/01457630701673170}
}

@article{Bing09Asundi,
  title = {Measurement of Coefficient of Thermal Expansion of Films Using Digital Image Correlation Method},
  author = {Bing, Pan and {Hui-min}, Xie and Tao, Hua and Asundi, Anand},
  year = 2009,
  journal = {Polymer Testing},
  volume = {28},
  number = {1},
  pages = {75--83},
  issn = {0142-9418},
  doi = {10.1016/j.polymertesting.2008.11.004}
}

@article{Blanc13Joubert,
  title = {Patterning of Silicon Nitride for {{CMOS}} Gate Spacer Technology. {{I}}. {{Mechanisms}} Involved in the Silicon Consumption in {{CH3F}}/{{O2}}/{{He}} High Density Plasmas},
  author = {Blanc, Romuald and Leverd, Fran{\c c}ois and David, Thibaut and Joubert, Olivier},
  year = 2013,
  journal = {Journal of Vacuum Science \& Technology B},
  volume = {31},
  number = {5},
  pages = {051801},
  issn = {2166-2746},
  doi = {10.1116/1.4816466}
}

@article{Bontempi01Pirri,
  title = {X-Ray Reflectivity Spectra of Ultrathin Films and Nanometric Multilayers: {{Experiment}} and Simulation},
  shorttitle = {X-Ray Reflectivity Spectra of Ultrathin Films and Nanometric Multilayers},
  author = {Bontempi, E. and Depero, L. E. and Sangaletti, L. and Giorgis, F. and Pirri, C. F.},
  year = 2001,
  journal = {Journal of Materials Research},
  volume = {16},
  number = {9},
  pages = {2556--2561},
  issn = {2044-5326},
  doi = {10.1557/JMR.2001.0350},
  langid = {english}
}

@article{Chou20Drndic,
  title = {Lifetime and {{Stability}} of {{Silicon Nitride Nanopores}} and {{Nanopore Arrays}} for {{Ionic Measurements}}},
  author = {Chou, Yung-Chien and Masih Das, Paul and Monos, Dimitri S. and Drndi{\'c}, Marija},
  year = 2020,
  journal = {ACS Nano},
  volume = {14},
  number = {6},
  pages = {6715--6728},
  publisher = {American Chemical Society},
  issn = {1936-0851},
  doi = {10.1021/acsnano.9b09964}
}

@article{Chu67Gruber,
  title = {The {{Preparation}} and {{Properties}} of {{Amorphous Silicon Nitride Films}}},
  author = {Chu, T. L. and Lee, C. H. and Gruber, G. A.},
  year = 1967,
  journal = {Journal of The Electrochemical Society},
  volume = {114},
  number = {7},
  pages = {717},
  publisher = {IOP Publishing},
  issn = {1945-7111},
  doi = {10.1149/1.2426715},
  langid = {english}
}

@article{Chuang04Ghodssi,
  title = {Mechanical Property Characterization of {{LPCVD}} Silicon Nitride Thin Films at Cryogenic Temperatures},
  author = {Chuang, Wen-Hsien and Luger, T. and Fettig, R.K. and Ghodssi, R.},
  year = 2004,
  journal = {Journal of Microelectromechanical Systems},
  volume = {13},
  number = {5},
  pages = {870--879},
  issn = {1941-0158},
  doi = {10.1109/JMEMS.2004.836815},
  langid = {american}
}

@article{Ciftlik11Gijs,
  title = {Parylene to Silicon Nitride Bonding for Post-Integration of High Pressure Microfluidics to {{CMOS}} Devices},
  author = {Ciftlik, Ata Tuna and Gijs, Martin A. M.},
  year = 2011,
  journal = {Lab on a Chip},
  volume = {12},
  number = {2},
  pages = {396--400},
  publisher = {The Royal Society of Chemistry},
  issn = {1473-0189},
  doi = {10.1039/C1LC20727J},
  langid = {english}
}

@article{Dekker07,
  title = {Solid-State Nanopores},
  author = {Dekker, Cees},
  year = 2007,
  journal = {Nature Nanotechnology},
  volume = {2},
  number = {4},
  pages = {209--215},
  publisher = {Nature Publishing Group},
  issn = {1748-3395},
  doi = {10.1038/nnano.2007.27},
  copyright = {2007 Springer Nature Limited},
  langid = {english}
}

@article{Dudescu06Nebel,
  title = {Characterisation of {{Thermal Expansion Coefficient}} of {{Anisotropic Materials}} by {{Electronic Speckle Pattern Interferometry}}},
  author = {Dudescu, C. and Naumann, J. and Stockmann, M. and Nebel, S.},
  year = 2006,
  journal = {Strain},
  volume = {42},
  number = {3},
  pages = {197--205},
  issn = {1475-1305},
  doi = {10.1111/j.1475-1305.2006.00271.x},
  langid = {english}
}

@article{Edwards04Sharpe,
  title = {Comparison of Tensile and Bulge Tests for Thin-Film Silicon Nitride},
  author = {Edwards, R. L. and Coles, G. and Sharpe, W. N.},
  year = 2004,
  journal = {Experimental Mechanics},
  volume = {44},
  number = {1},
  pages = {49--54},
  issn = {1741-2765},
  doi = {10.1007/BF02427976},
  langid = {english}
}

@article{Han23Jiang,
  title = {Advances in High-Performance {{MEMS}} Pressure Sensors: Design, Fabrication, and Packaging},
  shorttitle = {Advances in High-Performance {{MEMS}} Pressure Sensors},
  author = {Han, Xiangguang and Huang, Mimi and Wu, Zutang and Gao, Yi and Xia, Yong and Yang, Ping and Fan, Shu and Lu, Xuhao and Yang, Xiaokai and Liang, Lin and Su, Wenbi and Wang, Lu and Cui, Zeyu and Zhao, Yihe and Li, Zhikang and Zhao, Libo and Jiang, Zhuangde},
  year = 2023,
  journal = {Microsystems \& Nanoengineering},
  volume = {9},
  number = {1},
  pages = {156},
  publisher = {Nature Publishing Group},
  issn = {2055-7434},
  doi = {10.1038/s41378-023-00620-1},
  copyright = {2023 The Author(s)},
  langid = {english}
}

@book{Haynes17,
  title = {{{CRC Handbook}} of {{Chemistry}} and {{Physics}}, 97th {{Edition}}},
  editor = {Haynes, William M.},
  year = 2017,
  publisher = {CRC Press},
  address = {Boca Raton London New York},
  isbn = {978-1-4987-5428-6},
  langid = {english}
}

@article{Hirai81Goto,
  title = {Density and Deposition Rates of Amorphous {{CVD-Si3N4}} Including Carbon},
  author = {Hirai, Toshio and Goto, Takashi},
  year = 1981,
  journal = {Journal of Materials Science},
  volume = {16},
  number = {10},
  pages = {2877--2882},
  issn = {1573-4803},
  doi = {10.1007/BF00552973},
  langid = {english}
}

@article{James01Evans,
  title = {A Review of Measurement Techniques for the Thermal Expansion Coefficient of Metals and Alloys at Elevated Temperatures},
  author = {James, J. D. and Spittle, J. A. and Brown, S. G. R. and Evans, R. W.},
  year = 2001,
  journal = {Measurement Science and Technology},
  volume = {12},
  number = {3},
  pages = {R1},
  issn = {0957-0233},
  doi = {10.1088/0957-0233/12/3/201},
  langid = {english}
}

@article{Li26Cui1,
  title = {Cryogenic Thermometry via Nanofilm Thermal-Strain Optomechanics},
  author = {Li, Guangyu and Wu, Hui and Zheng, Weidong and Cao, Qun and He, Zhe and Cui, Zheng},
  year = 2026,
  journal = {Photonics Research},
  volume = {14},
  number = {8},
  pages = {3318--3325},
  publisher = {Optica Publishing Group},
  issn = {2327-9125},
  doi = {10.1364/PRJ.590925},
  copyright = {\copyright{} 2026 Chinese Laser Press},
  langid = {english}
}

@article{Li98Kojima,
  title = {X-Ray Photoelectron Spectroscopy and Grazing Incidence {{X-ray}} Reflectivity Study of Silicon Nitride Thin Films},
  author = {Li, Boquan and Fujimoto, Toshiyuki and Fukumoto, Natsuo and Honda, Kumasa and Kojima, Isao},
  year = 1998,
  journal = {Thin Solid Films},
  volume = {334},
  number = {1},
  pages = {140--144},
  issn = {0040-6090},
  doi = {10.1016/S0040-6090(98)01132-8}
}

@article{Liao18Goorsky,
  title = {Coefficients of Thermal Expansion of Single Crystalline {$\beta$}-{{Ga}}{\textsubscript{2}}{{O}}{\textsubscript{3}} and in-Plane Thermal Strain Calculations of Various Materials Combinations with {$\beta$}-{{Ga}}{\textsubscript{2}}{{O}}{\textsubscript{3}}},
  author = {Liao, Michael E. and Li, Chao and Yu, Hsuan Ming and Rosker, Eva and Tadjer, Marko J. and Hobart, Karl D. and Goorsky, Mark S.},
  year = 2018,
  journal = {APL Materials},
  volume = {7},
  number = {2},
  pages = {022517},
  issn = {2166-532X},
  doi = {10.1063/1.5054327}
}

@article{Lindsay-Scott07Dobson,
  title = {Thermal Expansion of {{CaIrO}}{\textsubscript{3}} Determined by {{X-ray}} Powder Diffraction},
  author = {{Lindsay-Scott}, Alex and Wood, Ian G. and Dobson, David P.},
  year = 2007,
  journal = {Physics of the Earth and Planetary Interiors},
  volume = {162},
  number = {1},
  pages = {140--148},
  issn = {0031-9201},
  doi = {10.1016/j.pepi.2007.04.003}
}

@article{Martyniuk06Faraone,
  title = {Environmental Stability and Cryogenic Thermal Cycling of Low-Temperature Plasma-Deposited Silicon Nitride Thin Films},
  author = {Martyniuk, M. and Antoszewski, J. and Musca, C. A. and Dell, J. M. and Faraone, L.},
  year = 2006,
  journal = {Journal of Applied Physics},
  volume = {99},
  number = {5},
  pages = {053519},
  issn = {0021-8979},
  doi = {10.1063/1.2179969}
}

@phdthesis{Martyniuk06THESIS,
  title = {Low-Temperature Micro-Opto-Electro-Mechanical Technologies for Temperature Sensitive Substrates},
  author = {Martyniuk, Mariusz},
  year = 2006,
  langid = {american},
  school = {University of Western Australia}
}

@article{Middelmann15Schodel,
  title = {Thermal Expansion Coefficient of Single-Crystal Silicon from 7 {{K}} to 293 {{K}}},
  author = {Middelmann, Thomas and Walkov, Alexander and Bartl, Guido and Sch{\"o}del, Ren{\'e}},
  year = 2015,
  journal = {Physical Review B},
  volume = {92},
  number = {17},
  pages = {174113},
  publisher = {American Physical Society},
  doi = {10.1103/PhysRevB.92.174113}
}

@book{Pierson99,
  title = {Handbook of {{Chemical Vapor Deposition}}},
  author = {Pierson, Hugh O.},
  year = 1999,
  publisher = {Noyes Publications},
  address = {New York},
  doi = {10.1016/C2009-0-20448-X},
  isbn = {978-0-8155-1300-1},
  langid = {english}
}

@article{Pu20Robertson,
  title = {Liquid Cell Transmission Electron Microscopy and Its Applications},
  author = {Pu, Shengda and Gong, Chen and Robertson, Alex W.},
  year = 2020,
  journal = {Royal Society Open Science},
  volume = {7},
  number = {1},
  pages = {191204},
  issn = {2054-5703},
  doi = {10.1098/rsos.191204}
}

@article{Retajczyk80Sinha,
  title = {Elastic Stiffness and Thermal Expansion Coefficients of Various Refractory Silicides and Silicon Nitride Films},
  author = {Retajczyk, T. F. and Sinha, A. K.},
  year = 1980,
  journal = {Thin Solid Films},
  volume = {70},
  number = {2},
  pages = {241--247},
  issn = {0040-6090},
  doi = {10.1016/0040-6090(80)90364-8}
}

@article{Retajczyk80Sinha1,
  title = {Elastic Stiffness and Thermal Expansion Coefficient of {{BN}} Films},
  author = {Retajczyk, Jr., T. F. and Sinha, A. K.},
  year = 1980,
  journal = {Applied Physics Letters},
  volume = {36},
  number = {2},
  pages = {161--163},
  issn = {0003-6951},
  doi = {10.1063/1.91415}
}

@article{Santucci01Lozzi,
  title = {X-Ray Reflectivity Studies of Very Thin Films of Silicon Oxide and Silicon Oxide--Silicon Nitride Stacked Structures},
  author = {Santucci, S. and {la Cecilia}, A. V. and DiGiacomo, A. and Phani, R. A. and Lozzi, L.},
  year = 2001,
  journal = {Journal of Non-Crystalline Solids},
  series = {3rd {{Symp}}. on {{SiO2}} and {{Advanced Dielectrics}}},
  volume = {280},
  number = {1},
  pages = {228--234},
  issn = {0022-3093},
  doi = {10.1016/S0022-3093(00)00380-X}
}

@article{Serra16Bonaldi,
  title = {Microfabrication of Large-Area Circular High-Stress Silicon Nitride Membranes for Optomechanical Applications},
  author = {Serra, E. and Bawaj, M. and Borrielli, A. and Di Giuseppe, G. and Forte, S. and Kralj, N. and Malossi, N. and Marconi, L. and Marin, F. and Marino, F. and Morana, B. and Natali, R. and Pandraud, G. and Pontin, A. and Prodi, G. A. and Rossi, M. and Sarro, P. M. and Vitali, D. and Bonaldi, M.},
  year = 2016,
  journal = {AIP Advances},
  volume = {6},
  number = {6},
  pages = {065004},
  issn = {2158-3226},
  doi = {10.1063/1.4953805}
}

@article{Stoney09,
  title = {The Tension of Metallic Films Deposited by Electrolysis},
  author = {Stoney, George Gerald},
  year = 1909,
  journal = {Proceedings of the Royal Society of London. Series A, Containing Papers of a Mathematical and Physical Character},
  volume = {82},
  number = {553},
  pages = {172--175},
  issn = {0950-1207},
  doi = {10.1098/rspa.1909.0021}
}

@article{Thurn04Hughey,
  title = {Evaluation of Film Biaxial Modulus and Coefficient of Thermal Expansion from Thermoelastic Film Stress Measurements},
  author = {Thurn, Jeremy and Hughey, Michael P.},
  year = 2004,
  journal = {Journal of Applied Physics},
  volume = {95},
  number = {12},
  pages = {7892--7897},
  issn = {0021-8979},
  doi = {10.1063/1.1753662}
}

@article{Tiilikainen08Lipsanen,
  title = {Novel Method for Error Limit Determination in X-Ray Reflectivity Analysis},
  author = {Tiilikainen, J and Mattila, M and Hakkarainen, T and Lipsanen, H},
  year = 2008,
  journal = {Journal of Physics D: Applied Physics},
  volume = {41},
  number = {11},
  pages = {115302},
  issn = {0022-3727},
  doi = {10.1088/0022-3727/41/11/115302},
  langid = {english}
}

@article{Vavrille24Estevez,
  title = {Advances in the {{Thermal Study}} of {{Polymers}} for {{Microelectronics Using}} the {{Thermally Induced Curvature Approach}}},
  author = {Vavrille, Benjamin and Vignoud, Lionel and Chapelon, Laurent-Luc and Estevez, Rafael},
  year = 2024,
  journal = {IEEE Transactions on Semiconductor Manufacturing},
  volume = {37},
  number = {3},
  pages = {251--259},
  issn = {1558-2345},
  doi = {10.1109/TSM.2024.3410513}
}

@incollection{Wolff99,
  title = {Thermal {{Expansion}} of {{Thin Films}}: {{A Review}}},
  shorttitle = {Thermal {{Expansion}} of {{Thin Films}}},
  booktitle = {Thermal Conductivity},
  author = {Wolff, Emest G.},
  year = 1999,
  volume = {24},
  pages = {368--387},
  publisher = {CRC Press},
  isbn = {978-1-56676-711-8}
}

@article{Zhou95Chen,
  title = {Theoretical Foundation of {{X-ray}} and Neutron Reflectometry},
  author = {Zhou, Xiao-Lin and Chen, Sow-Hsin},
  year = 1995,
  journal = {Physics Reports},
  volume = {257},
  number = {4},
  pages = {223--348},
  issn = {0370-1573},
  doi = {10.1016/0370-1573(94)00110-O}
}

@article{Zwickl08Harris,
  title = {High Quality Mechanical and Optical Properties of Commercial Silicon Nitride Membranes},
  author = {Zwickl, B. M. and Shanks, W. E. and Jayich, A. M. and Yang, C. and Bleszynski Jayich, A. C. and Thompson, J. D. and Harris, J. G. E.},
  year = 2008,
  journal = {Applied Physics Letters},
  volume = {92},
  number = {10},
  pages = {103125},
  issn = {0003-6951},
  doi = {10.1063/1.2884191}
}

\end{document}